# Structural order promotes efficient separation of delocalized charges at molecular heterojunctions

Xiangkun Jia,† Lorenzo Soprani,† Giacomo Londi,† Seyed Mehrdad Hosseini, Felix Talnack, Stefan Mannsfeld, Safa Shoaee, Dieter Neher, Sebastian Reineke, Luca Muccioli, Gabriele D'Avino, Koen Vandewal,* David Beljonne,* and Donato Spoltore*

(†: these three authors contributed equally. *: corresponding authors.)

X. Jia, S. Reineke, D. Spoltore

Dresden Integrated Center for Applied Physics and Photonic Materials (IAPP) and Institute for Applied Physics, Technische Universität Dresden, 01187 Dresden, Germany

X. Jia

Present address: Department of Chemical and Biomolecular Engineering, National University of Singapore, 4 Engineering Drive 4, Singapore 117585, Singapore

L. Soprani, L. Muccioli

Department of Industrial Chemistry "Toso Montanari", University of Bologna, 40136 Bologna, Italy

G. Londi, D. Beljonne

Laboratory for Chemistry of Novel Materials, University of Mons, B-7000 Mons, Belgium

E-mail: david.beljonne@umons.ac.be

S. M. Hosseini, S. Shoaee, D. Neher

Institute of Physics and Astronomy, University of Potsdam, Karl-Liebknecht-Str. 24-25, 14476 Potsdam, Germany

F. Talnack, S. Mannsfeld

Center for Advancing Electronics Dresden (cfaed) and Faculty of Electrical and Computer Engineering, Technische Universität Dresden, 01062 Dresden, Germany

G. D'Avino

Grenoble Alpes University, CNRS, Grenoble INP, Institut Néel, 25 rue des Martyrs, 38042 Grenoble, France

K. Vandewal, D. Spoltore

Institute for Materials Research (IMO-IMOMEC), Hasselt University, Wetenschapspark 1, 3590 Diepenbeek, Belgium
Email: koen.vandewal@uhasselt.be

D. Spoltore

Department of Mathematical, Physical and Computer Sciences, University of Parma, V.le delle Scienze 7/A, 43124 Parma, Italy
Email: donato.spoltore@unipr.it

**Abstract**

The energetic landscape at the interface between electron donating and accepting molecular materials favors efficient conversion of intermolecular charge-transfer states (CTS) into free charge carriers in high-performance organic solar cells. Here, we elucidate how interfacial energetics, charge generation and radiative recombination are affected by structural ordering. We experimentally determine the CTS binding energy of a series of model, small molecule donor-acceptor blends, where the used acceptors (B2PYMPM, B3PYMPM and B4PYMPM) differ only in the nitrogen position of their lateral pyridine rings. We find that the formation of an ordered, face-on molecular packing in B4PYMPM is beneficial to efficient, field-independent charge separation, leading to fill factors over 70% in photovoltaic devices. This is rationalized by a comprehensive computational protocol showing that, compared to the more amorphous and isotropically oriented B2PYMPM, the higher order of the B4PYMPM molecules provides more delocalized CTS. Furthermore, we find no correlation between the quantum efficiency of radiative free charge carrier recombination and the bound or unbound nature of the CTS. This work highlights the importance of structural ordering at donor-acceptor interfaces for efficient free carrier generation and shows that more ordering and less bound CT states do not preclude efficient radiative recombination.

## 1. Introduction

Blends of electron donating (D) and electron accepting (A) organic compounds find their applications in organic solar cells (OSCs) and organic light emitting diodes (OLEDs). The power conversion efficiencies (PCEs) of OSCs have exceeded 18%,[1] due to the development of new strongly absorbing non-fullerene acceptors. Furthermore, D-A blends are regularly used as host materials for OLEDs with low driving voltage and improved power efficiency.[2] Indeed, electroluminescence quantum efficiencies of 21.7% have been reported for D-A (or exciplex) OLEDs.[3,4] Recently, some of those visible-light emitting D-A blends have been discovered to exhibit also a high photovoltaic quantum efficiency.[5] These particular systems, as compared to the more commonly studied photovoltaic blends with optical gaps in the near-infrared, benefit from reduced non-radiative decay rates for their visible light emitting charge-transfer (CT) states, but the origin of their efficient CT state dissociation has not been disclosed to date. Because of low dielectric constants in organic materials, both the photo-induced local exciton (LE) and the CT state are expected to be strongly bound by Coulomb forces.[6] This has raised a long-standing fundamental question: how can thermalized CT states overcome this binding energy and split into free charge carriers? Several mechanisms have been proposed to rationalize efficient charge generation in OSCs, such as interface morphology,[7,8] energetic disorder,[9,10] electrostatic effects,[11] charge delocalization,[12–15] and entropic considerations[16,17] but a consensus is still lacking.

In this work, we study vacuum deposited small molecule D-A blends with relevance for organic photovoltaics and light emission, consisting of several donors mixed with the acceptors B2PYMPM, B3PYMPM and B4PYMPM,[18] which only differ in the nitrogen position of substituted pyridine rings. These materials show a different preferential molecular orientation and structural order when forming a solid film, and diodes based on these blends have strongly varying photovoltaic and electroluminescence characteristics. Through the analysis of the CT absorption and emission spectra and of the temperature dependence of the open-circuit voltage ($V_{OC}$), we determine whether the CT states are bound and, if so, determine their CT state binding energy ($E_B$).[5] In B3PYMPM and B2PYMPM-based blends with poor photovoltaic performance, we find a sizable $E_B$ (~0.1 eV) and a CT state dissociation strongly depending on the electric field, while in B4PYMPM containing blends we find efficient, field-independent dissociation and fill factors (FFs) exceeding 70%. 2D grazing-incidence wide-angle X-ray scattering (GIWAXS) data reveal a relation between ordered, face-on molecular stacking and efficient CT state dissociation. Computational modelling, which combines molecular dynamics (MD) simulations, density functional theory (DFT) and microelectrostatic (ME) calculations as well as a tight-binding (TB) model designed to account for charge delocalization, suggests that more ordered B4PYMPM molecules offer more delocalized states for charge separation. Surprisingly, we find that the BF-DPB:B4PYMPM blend with unbound CT states and efficient charge carrier generation has the highest electroluminescence quantum efficiency of 1.5% in all studied blends. This indicates that emissive CT states are not necessarily strongly bound, which is encouraging for highly efficient organic photovoltaics with low non-radiative losses.

## 2. Chemical Structures and Photovoltaic Performance

The electron acceptor molecules B2PYMPM, B3PYMPM, and B4PYMPM differ in the position of the nitrogen atom in the pyridine rings (see **Figure 2a**). This small change in chemical structure drives strikingly large differences in molecular packing and electron mobility.[19,20] Four different small-molecular donors, being BF-DPB, NPB, NDDP, and m-MTDATA, were combined with the three acceptors. Their chemical structure, energy levels, and device architecture are shown in Figure S1 (Supporting Information). We characterized the photovoltaic and electroluminescence performance of these bulk-heterojunction (BHJ) D:A combinations, and their parameters are summarized in Table S1. Because of the large optical gap of the selected molecules, the main absorption peaks of photovoltaic external quantum efficiency curves (EQE$_{PV}$, Figure S2) are in the range of 300-400 nm, resulting in rather low photocurrents under AM 1.5G illumination. The $V_{OC}$s of B4PYMPM-containing devices are ~0.2 V lower than those obtained with either B2PYMPM or B3PYMPM because of their decreased $E_{CT}$ as a consequence of the relatively lower LUMO energy of the former (Figure S1).[18] However, the most intriguing finding, as shown in Figure 1b-d, is that regardless of the nature of the donor, B4PYMPM-based devices consistently produce very high FFs (> ~70%); in contrast, FFs in B2PYMPM- or B3PYMPM-based devices are poor (~ 20-40%). Low FFs can be a consequence of either poor charge transport or a poor and field-dependent dissociation of the CT states into free charge carriers. The time-delayed collection field (TDCF) measurements (see Figure1e-g and discussion below) reveal that the latter process dominates in B2PYMPM and B3PYMPM devices, and that conversely, efficient and nearly field-independent charge generation upon photoexcitation occurs in B4PYMPM systems.

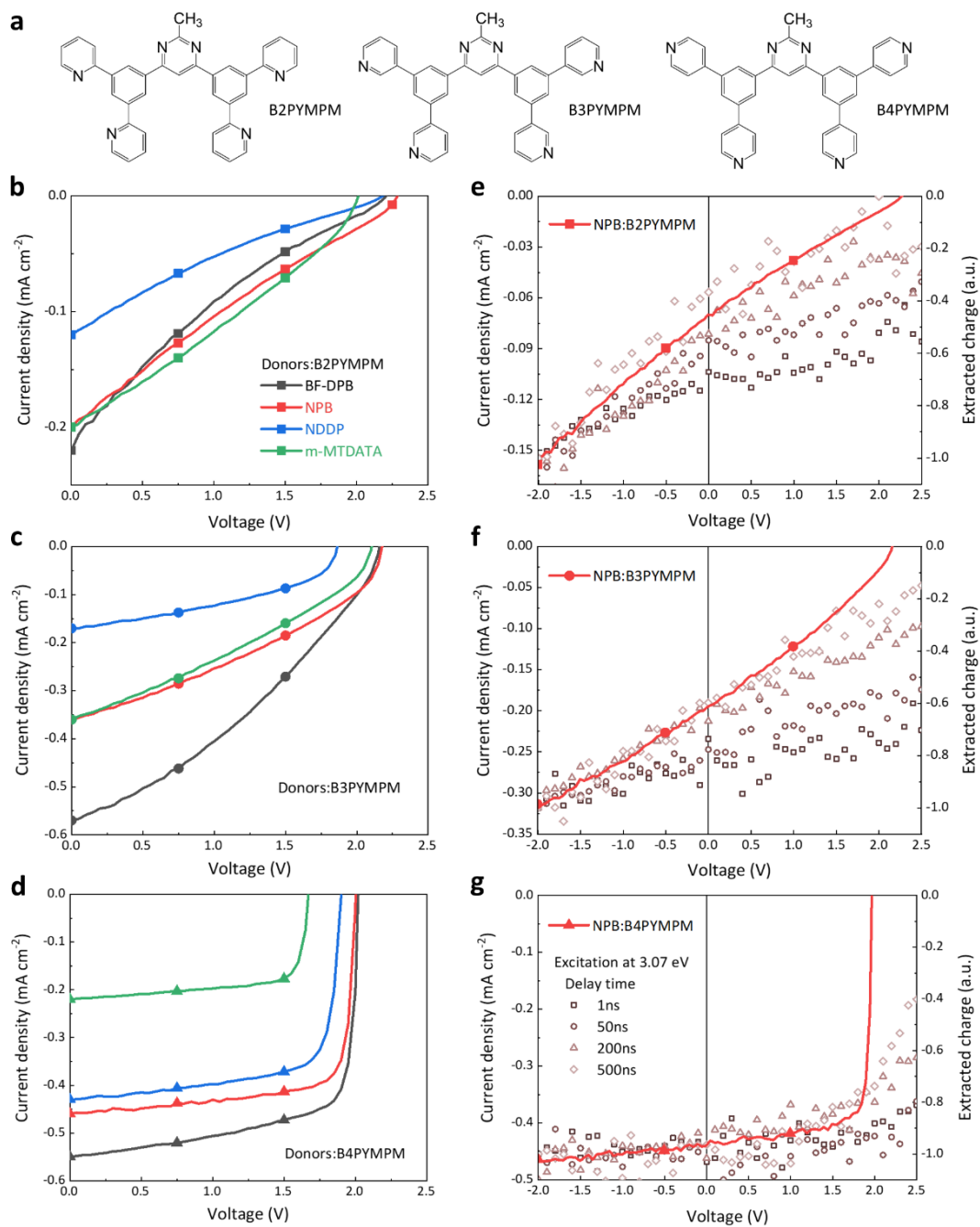

**Figure 1.** Chemical structures, photovoltaic performance, and time-delayed collection field (TDCF) measurements. a) Chemical structures of three investigated acceptor molecules: B2PYMPM, B3PYMPM, and B4PYMPM. b-d) Current-density voltage characteristics measured under simulated AM 1.5G solar illumination for devices based on three acceptors combined with four different donors. All these standard devices have a 30 nm thick active layer. Black, red, blue, and green colors are chosen for representing donor molecule BF-DPB, NPB, NDDP and m-MTADATA, respectively, while filled square, circle, and triangle denote B2PYMPM, B3PYMPM, and B4PYMPM acceptors, respectively. e-g) Current density-voltage curves are represented as solid lines for NPB:B2PYMPM, NPB:B3PYMPM, and NPB:B4PYMPM. Devices for TDCF measurements have a 100 nm thick active layer. The relative number of generated charge carriers extracted in the TDCF measurement as a function of applied bias is shown on the right axis. A short laser pulse (~4 ns) with a wavelength of 404 nm (3.07 eV) is used to excite samples, and the collection voltage is -3 V with a delay time in the range of 1-500 ns.

## 3. Time-Delayed Collection Field (TDCF) Measurements

Since high FFs were found for all B4PYMPM based systems, we focus on NPB-based devices, which can achieve a FF of 81% even with a 100 nm thick active layer (EQE$_{PV}$ and J-V curves of the 100 nm device in Figure S3). This is convenient because active layers thicker than the usual 30 nm are required for TDCF measurements to minimize the capacitance and the corresponding RC rise time of the device. Shown in Figure 1e-g is the total charge extracted from the device for different delay times between the center of the laser pulse and a strong reverse collection voltage ($V_{coll}$), as a function of applied bias. When $V_{coll}$ is given 1 ns after pulsed illumination, the amount of free charges depends strongly on applied voltage for the NPB:B2PYMPM and NPB:B3PYMPM combinations. This indicates a geminate recombination process occurring on a ns timescale.[21,22] Increasing the delay time up to 500 ns causes the field-dependence of the extracted charge to increase, which is the result of non-geminate recombination starting to take place at these timescales.[23–25] As a result of the strongly field-dependent charge generation and rapid non-geminate recombination in B2PYMPM and B3PYMPM system, the FF is severely reduced. In contrast, a much weaker to no field dependent generation is observed in B4PYMPM systems in Figure 1g at timescales between 1-500 ns, suggesting much reduced geminate and non-geminate recombination. It was recently shown that both recombination processes are interlinked via the efficiency of CT dissociation.[26] This is because non-geminate recombination involves the reformation of the CT states, which in case of efficient dissociation reduces the effective decay rate of free charge carriers. As a result, the FF is significantly improved.

## 4. Binding Energy of Charge-Transfer State

Information on the relative energy of the CT state and free charge carriers can be obtained via a combination of sensitive optical spectroscopy and temperature-dependent $V_{OC}$ measurements. As shown in **Figure 2a-c**, sensitive electroluminescence (EL) and EQE$_{PV}$ measurements were performed to determine the energy of CT states ($E_{CT}$). For the BF-DPB:B2PYMPM combination, the $E_{CT}$ value is 2.71 eV, which is larger than those of BF-DPB:B3PYMPM ($E_{CT}$ = 2.57 eV) and BF-DPB:B4PYMPM ($E_{CT}$ = 2.49 eV) combinations, following the same trend as the corresponding $V_{OC}$ values (Table S2). In order to determine whether or not these CT states are bound, the $E_{CT}$ values of these blends are compared with $E_0$, whose values are determined as in Figure 2d-f from the extrapolation of temperature-dependent suns-$V_{OC}$ measurements to 0 K.[5] This energy $E_0$ corresponds well to the activation energy of electroluminescence (Figure S4). We notice here that, due to the increase of the LUMO energy of B2PYMPM, $E_{CT}$ becomes larger and approaches the optical gap of BF-DPB. As a result, a relatively larger injection current is needed to obtain a clear spectrum of the CT emission.

When CT state dissociation into free charge carriers is much faster than CT state decay, $E_0$ is expected to be equal to $E_{CT}$ (more details in appendix 1 of SI, and ref [27]). While it might still have a finite binding energy, the CT state behaves as it would be unbound as this binding energy is quickly overcome, within the lifetime of the CT state. If, on the other hand, CT state dissociation is slower or comparable to CT state decay, $E_0$ will be higher than $E_{CT}$, and will approximate the energy of an unbound electron-hole (e-h) pair.[27] In this case, the CT state is bound and its binding energy can be estimated by $E_B = E_0 - E_{CT}$.

For the BF-DPB:B4PYMPM system we find that $E_{CT}$ is equal to $E_0$, which indicates[27] that CT state dissociation is much faster than CT state decay. Even upon reformation by encounter of positive and negative charge carriers, there is a high probability for CT state dissociation, resulting in an equilibrium between free charge carriers and CT states before decay. Indeed, for this material blend, efficient and field independent dissociation, FFs > 70% and photovoltaic internal quantum efficiency $IQE_{PV}$ > 82% were found.[5] We want to stress once more that this result does not necessarily imply that the CT state has no binding energy: $E_{CT}$ being equal to $E_0$ means that free charge carriers are quickly formed, before the CT state decays. In this case, the value of $E_0$-$E_{CT}$ approaching zero (or even being slightly below zero) does not necessarily correspond to the CT state binding energy.[27]

On the other hand, for B2PYMPM- and B3PYMPM-based blends, $E_0$ is significantly higher than $E_{CT}$. This indicates that CT states decay before dissociation is significant[27] and the CT state binding energy can be approximated by $E_B = E_0 - E_{CT}$ (appendix 1 in SI and ref [27]), giving ~110 meV for B2PYMPM and B3PYMPM based blends. In these blends, e-h separation is slowed down so that CT state decay becomes competitive, translating into a field-dependent free charge carrier generation and a low FF.

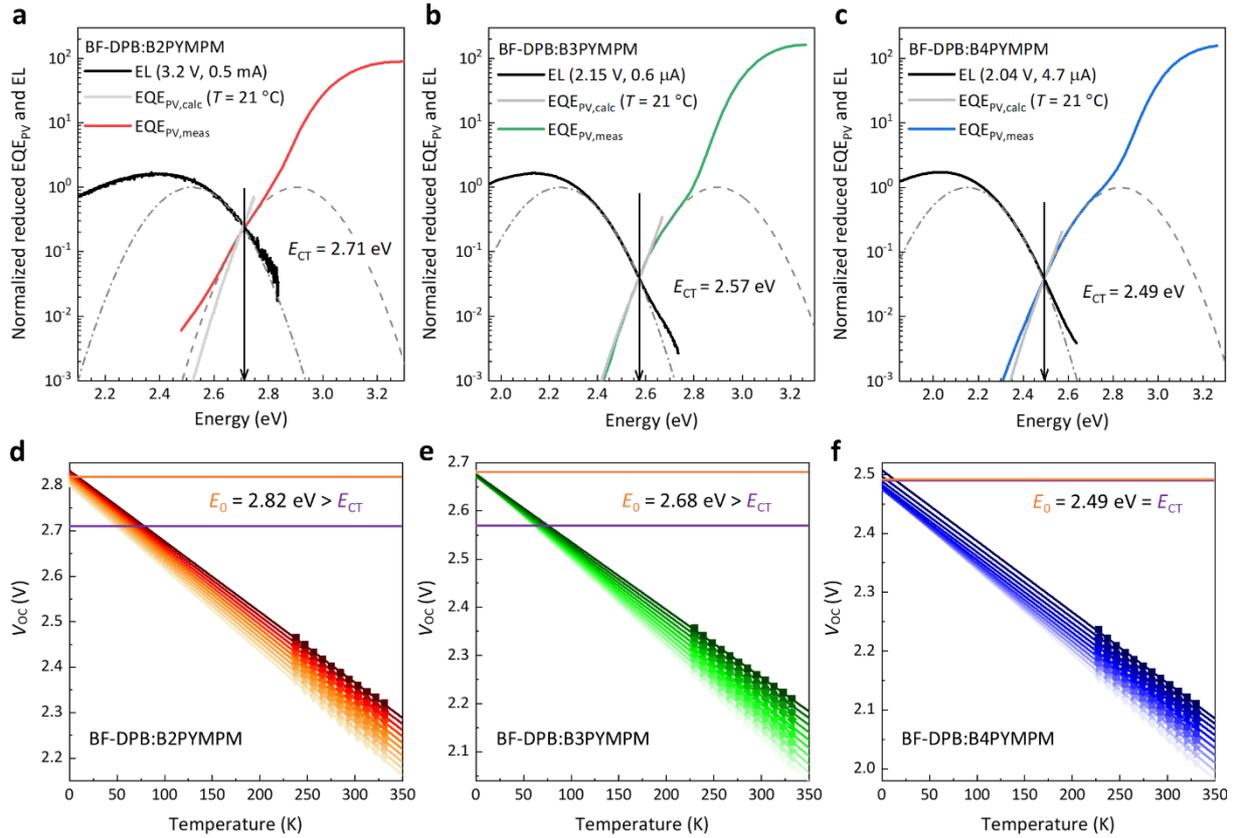

**Figure 2.** Sensitive EL and EQE$_{PV}$ measurements and determination of $E_0$. a-c) Normalized reduced electroluminescence (EL) and photovoltaic external quantum efficiency (EQE$_{PV}$) spectra as a function of the photon energy for three BF-DPB-based exemplary devices. The sensitive EL spectra are measured under low injection current condition to make sure that injected charge carriers reach thermal equilibrium before recombination. Gray EQE$_{PV}$ curves are calculated from EL spectra under the reciprocity assumption between absorption and emission, and show an agreement with sensitively measured EQE$_{PV}$ curves. The dashed curves are Gaussian fits of the EL and EQE$_{PV}$ curves following the methods described in a previous work.[28] The energy of charge-transfer states ($E_{CT}$) is obtained at the crossing point between appropriately scaled EL and EQE$_{PV}$ curves, highlighted by vertical black arrows.[28] d-f) Temperature-dependent suns-$V_{OC}$ measurements for corresponding exemplary devices, in which the temperature varies in the range of 223-333 K in steps of 10 K. For each intensity, the energy $E_0$ is obtained by extrapolating temperature to 0 K. For the energy values with fitting errors smaller than 15 meV, the average value is taken and represented as $E_0$, shown in a horizontal orange line.[5] The optically determined $E_{CT}$ is plotted as a purple line. In the case that $E_0$ is significantly higher than $E_{CT}$, the binding energy of the CT state ($E_B$) is the energy difference between $E_0$ and $E_{CT}$: $E_B = E_0 - E_{CT}$.

This trend can be deduced and established not only for the BF-DPB systems, but also for the other donor materials BHJ systems (Figure S5-S7 and Table S3). The same D:A combinations were also fabricated in a planar heterojunction (PHJ) structure and investigated as well. Their photovoltaic performance and parameters are shown in Figure S8 and Table S4. In comparison to their BHJ counterparts, most of PHJ devices exhibit higher FFs, which may be attributed to their lower non-geminate recombination rates associated to planar structures.[29] However, we still measure considerable CT binding energies for the B2PYMPM- and B3PYMPM-based systems (see Figure S9-S10 and Table S5). This is consistent with what we found in BHJ devices. All these results support the view that CT states in B4PYMPM-containing systems dissociate much faster than they decay, while for B2PYMPM- and B3PYMPM-based systems an $E_B$ greater than 110 meV prevents Coulombically bound e-h pairs from fast dissociation.

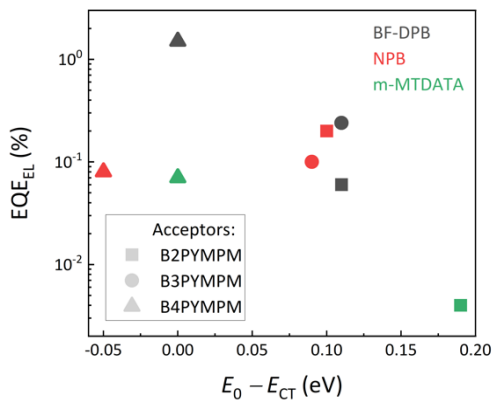

**Figure 3.** The relation between EQE$_{EL}$ and $E_0 - E_{CT}$. The symbols with black, red, and green colors represent BF-DPB-, NPB-, and m-MTDATA-based devices, respectively. Filled squares, circles, and triangles denote acceptor molecules B2PYMPM, B3PYMPM, and B4PYMPM, respectively. The EQE$_{EL}$ values are determined during the sensitive EL measurement; the injection conditions are shown together with sensitive EL curves.

**Figure 3** shows the measured electroluminescence external quantum efficiencies (EQE$_{EL}$) of all investigated BHJ devices, plotted as a function of $E_0$-$E_{CT}$. We find that B4PYMPM-based systems, where $E_0$-$E_{CT}$ is around 0, have a comparable or even higher EQE$_{EL}$ as compared to the devices based on B2PYMPM and B3PYMPM. In particular, the EQE$_{EL}$ of BF-DPB:B4PYMPM blend reaches 1.5%, showing that fast dissociation of CT states in equilibrium with free carriers do not necessarily have lower emission efficiencies than strongly bound states.

## 5. Molecular Orientation and Packing

To determine a possible microstructural origin to more efficient CT state dissociation of B4PYMPM as compared to B3PYMPM and B2PYMPM based systems, we conducted grazing-incidence wide-angle X-ray scattering (GIWAXS) measurements on both neat acceptor films and BF-DPB:A blend films. The 2D GIWAXS patterns of B2PYMPM, B3PYMPM and B4PYMPM pristine films on silicon substrates are presented in **Figure 44a-c** and the corresponding population orientation analysis are shown in Figure 44d-f (where an angle of 90 degrees means completely lying-down alignment and an angle of 0 degrees represents a standing-up arrangement with respect to the substrate). In all three GIWAXS images a ring at the scattering vector $q$ of 1.7 Å$^{-1}$ on the $Q_z$ axis is observed, this diffraction pattern is attributed to the π-π stacking of molecules. To investigate the orientational order of the three different molecules, the intensity along this ring is integrated on an χ-arc from 0° to 80°. The intensity along this χ-arc is shown Figure 4d-f for B2PYMPM, B3PYMPM and B4PYMPM, respectively. The intensity is corrected by cos(χ) to directly correlate the observed intensity to the amount of material orientated under a certain angle. From this analysis we can deduce that B4PYMPM molecules are orientated 44% face-on in regard to the substrate, surpassing both B3PYMPM (30.8%) and B2PYMPM (12.2%). In fact, most of B2PYMPM molecules are isotropically oriented (62.6%), as shown in Figure 44d and Table S6. This shows the increased orientational order of B4PYMPM, as compared to B3PYMPM and B2PYMPM. The full width at half maximum of the ring in the $Q_z$ direction 1.7 Å$^{-1}$ is 0.46 Å$^{-1}$ for B4PYMPM and 0.62 Å$^{-1}$ for B2PYMPM. As the peak width and the crystallinity are inversely correlated, this shows that B4PYMPM has an improved crystalline order in comparison to B2PYMPM. The same holds true for the in-plane direction, which can be observed from the small diffraction feature located at $Q_{xy}$ = 1.1 Å$^{-1}$.

Compared to neat acceptor films, blends of the acceptors with BF-DPB show a relatively higher isotropic percentage that, together with the significant increase in the peak width, suggests more amorphous molecular orientations in the blend films, which is attributed to the introduction of the amorphous donor material BF-DPB (see Figure S11 and Table S7). However, B4PYMPM-based blend films still show the highest face-on ratio and the lowest isotropically orientated amount of material among blend films, which hints to a higher degree of structural order compared to B2PYMPM and B3PYMPM, even if to a lower extent compared to the neat acceptors films

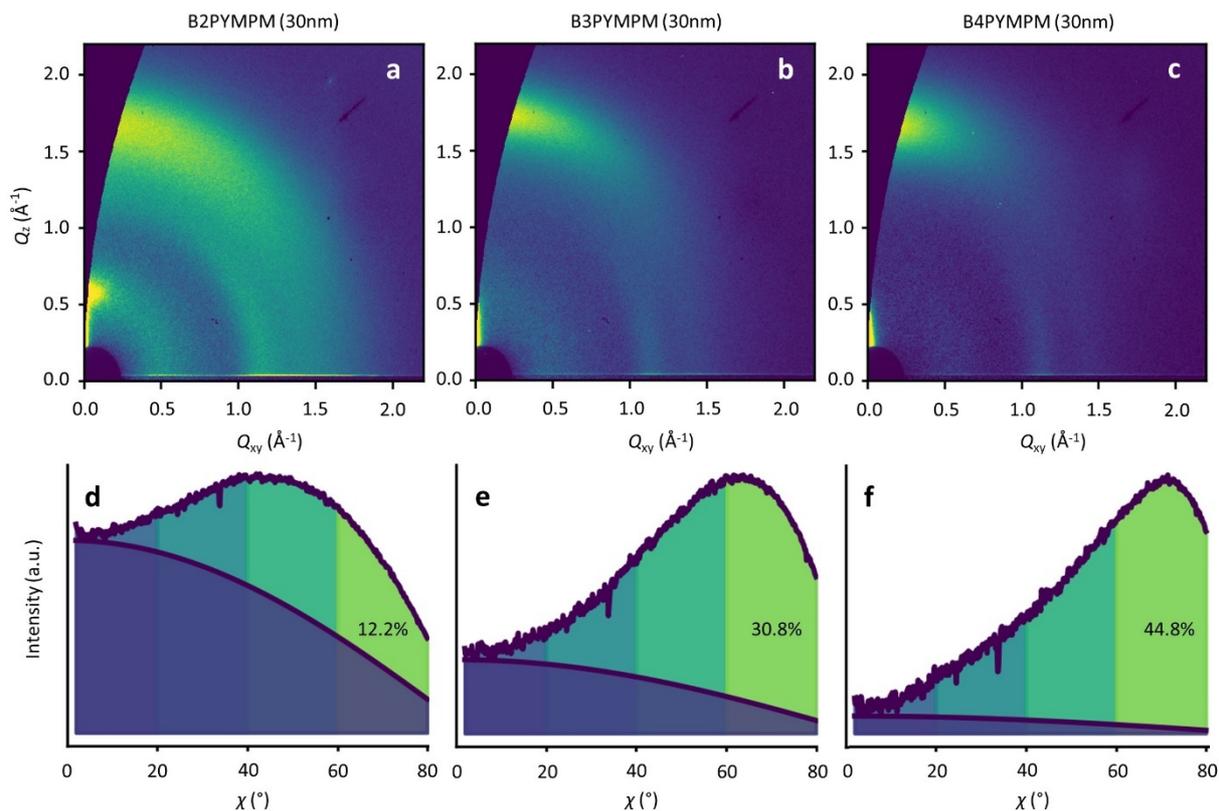

**Figure 4.** Morphology characterization. a-c) GIWAXS data of 30 nm thick neat B2PYMPM, B3PYMPM, and B4PYMPM deposited on silicon substrate, respectively. d-f) Corresponding population orientation analysis. The blue area visible at all angles denotes the intensity originating from isotropic orientated molecules. Angle χ equal to 0 degree indicates a fully edge-on molecular alignment, while a χ value of 90 degrees represents a completely face-on orientation for the π-π stacking diffraction.

## 6. Modelling the Interfacial Charge-Transfer Exciton Manifold

In an effort to understand the superior charge generation efficiency of B4PYMPM with respect to B3PYMPM and B2PYMPM, we set up and applied a multilevel computational approach that combines several techniques.[15,30] Charge delocalization effects on intermolecular e-h states were described with a tight binding (TB) model. The model was carefully parametrized starting from atomistic MD simulations for two D:A PHJ samples (**Figure 5a**): a BF-DPB:B2PYMPM junction between amorphous phases, and an interface between an amorphous BF-DPB phase and a crystalline B4PYMPM one. The choice of the acceptor phase morphology was guided by GIWAXS data for bilayer samples (see SI), which indicate a preferential face-on orientation of B4PYMPM molecules but a more isotropic one in the B2PYMPM film. For each MD sample, a 2D-periodic 8 nm-thick slab centered at the heterointerface was extracted from the MD morphology (Figure 5b) and employed in subsequent electronic structure calculations. DFT and microelectrostatic (ME) calculations were performed to assess the energy landscape of localized charge carriers for all the molecules at the interface, as well as the corresponding electron transfer integrals. This information was then utilized to construct a TB Hamiltonian for acceptor electron states that are Coulombically bound to a fixed hole in the donor phase. More in detail, for each BF-DPB at the interface, the TB model includes all the acceptor molecules within a hemisphere of radius $d$ (up to 6 nm) centered on the donor molecule being positively charged (Figure 5b,c). Repeating this procedure for the different hole positions and diagonalizing the corresponding Hamiltonian allows spanning the manifold of delocalized e-h configurations, including interfacial CT excitons and (partially) space-separated e-h states.

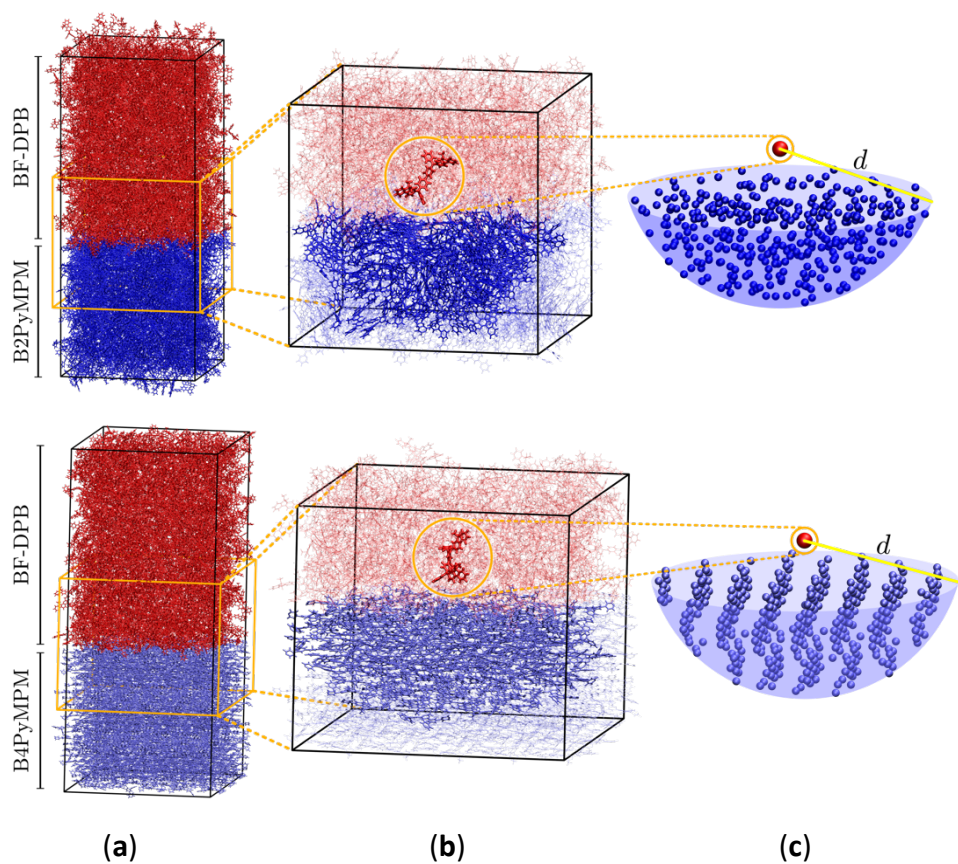

(a)  (b)  (c)

**Figure 5.** Illustration of the multiscale workflow for the modeling of electronic states at the BF-DPB:B2PYMPM (top) and BF-DPB:B4PYMPM (bottom) interfaces. (a) 2D periodic structure of planar heterojunctions obtained with MD simulation. Transfer integrals and charge transport energy levels have been computed for all the molecules within an (b) interfacial slab. A TB model describing electron states in the A phase was then set up for every fixed position of the hole on a given D molecule. The model includes all the A molecules within a radius $d$ from the hole. The BF-DPB D carrying the hole and the A molecules included is a radial selection are highlighted with thicker lines in (b). (c) Point-like molecule representation of such a selection, illustrating the more ordered structure of the B4PYMPM phase, as compared to the amorphous B2PYMPM.

The heat maps in Figure **6a-d** show the density of states (DOS, with probability rendered with a color code) as a function of energy and e-h distance. Two scenarios are considered: localized and delocalized electrons. Panels (a) and (c) depict the results obtained for holes and electrons localized over a single donor and a single acceptor molecule. That is obtained in the TB model by setting to zero all electron transfer integrals between acceptors. The plots in panels (b) and (d), instead, take into account electron delocalization, i.e. they are obtained from the energies of the eigenstates of a TB Hamiltonian in which the transfer integrals are calculated by extracting dimers from MD morphologies.

In Figure **6a-d** we also show the (i) arithmetic (solid blue lines) and (ii) Boltzmann averages (green lines) over states at a given e-h distance, superimposed to the DOS map. While (i) with its standard deviation (dashed blue lines) is a good estimator for the shape of the DOS, (ii) conveys information relevant to thermalized charge carrier at 300 K. Panels 6e-f display instead the participation ratio (PR), quantifying the number of molecules over which a given CT state is delocalized. Before discussing the consequences of these energy landscapes on charge separation, we remark that we did not observe any net trend in the energy levels with distance arising from molecular multipole moments at the interface. Such band bending effects, when present, can strongly impact the energetics of charge separation.[11,33] Similarly, we also excluded that entropy could assist charge separation in a significantly different way in the two systems (Figure S20), allowing us to focus on the potential energy profiles only.

From the arithmetic averages of the states near to the interface (Figure 6b and 6d, solid blue lines) we estimate that $E_{CT}$ amounts to ~3.7 eV and to ~3.3 eV for BF-DPB:B2PYMPM and DPB:B4PYMPM, respectively. This $E_{CT}$ difference of ~0.4 eV is in fair agreement with experimental measurements (Figure 2a,c). In both systems, owing to the absence of net electrostatic forces at the D:A interface, the $E_{CT}$ arithmetic averages follow roughly the same profile vs distance, consistent with the Coulomb interaction between charged donors and acceptors. It is hence more instructive to focus on Boltzmann-averaged profiles than on arithmetic averages. For B2PYMPM the profile barely changes including or not charge delocalization effects (solid and green lines in Figure 6a,b). This is due to the very small electron transfer integrals and the large energetic disorder in B2PYMPM, which both concur in leading to almost fully localized, *i.e.* PRs close to unity (see Figure 6e). Conversely, for the crystalline B4PYMPM sample, the low-energy states are delocalized over multiple acceptor units (PRs up to 3, see Figure 6f), with a corresponding widening of the DOS (see Figure 6d).

This different extent of charge delocalization has important consequences on the energetics of charge separation. We evaluated the barrier $E_B$ for charge separation from the plots in Figure 6b and 6d as the difference between the Boltzmann average energy (solid green lines) at the interface (~10 Å) and value assumed to represent an infinite distance (i.e. ~35 Å). In the B2PYMPM sample, interfacial CT excitons ($r_{eh}$ < 10 Å) have to overcome an $E_B$ of ~0.2 eV to move far away from the hole ($r_{eh}$ > 30 Å), irrespective of including or not delocalization effects. Conversely, a smaller barrier of ~0.1 eV is found for charge carriers in BF-DPB:B4PYMPM, as a result of the more pronounced charge delocalization in this system. This difference in charge delocalization between the two systems is in fair agreement with the experimental observations of a more strongly bound CT state for B2PYMPM based systems. We emphasize that such a difference is rooted in the difference of structural order between the two A phases, and it loosely depends on the actual orientation of molecules at the interface employed in the calculations.

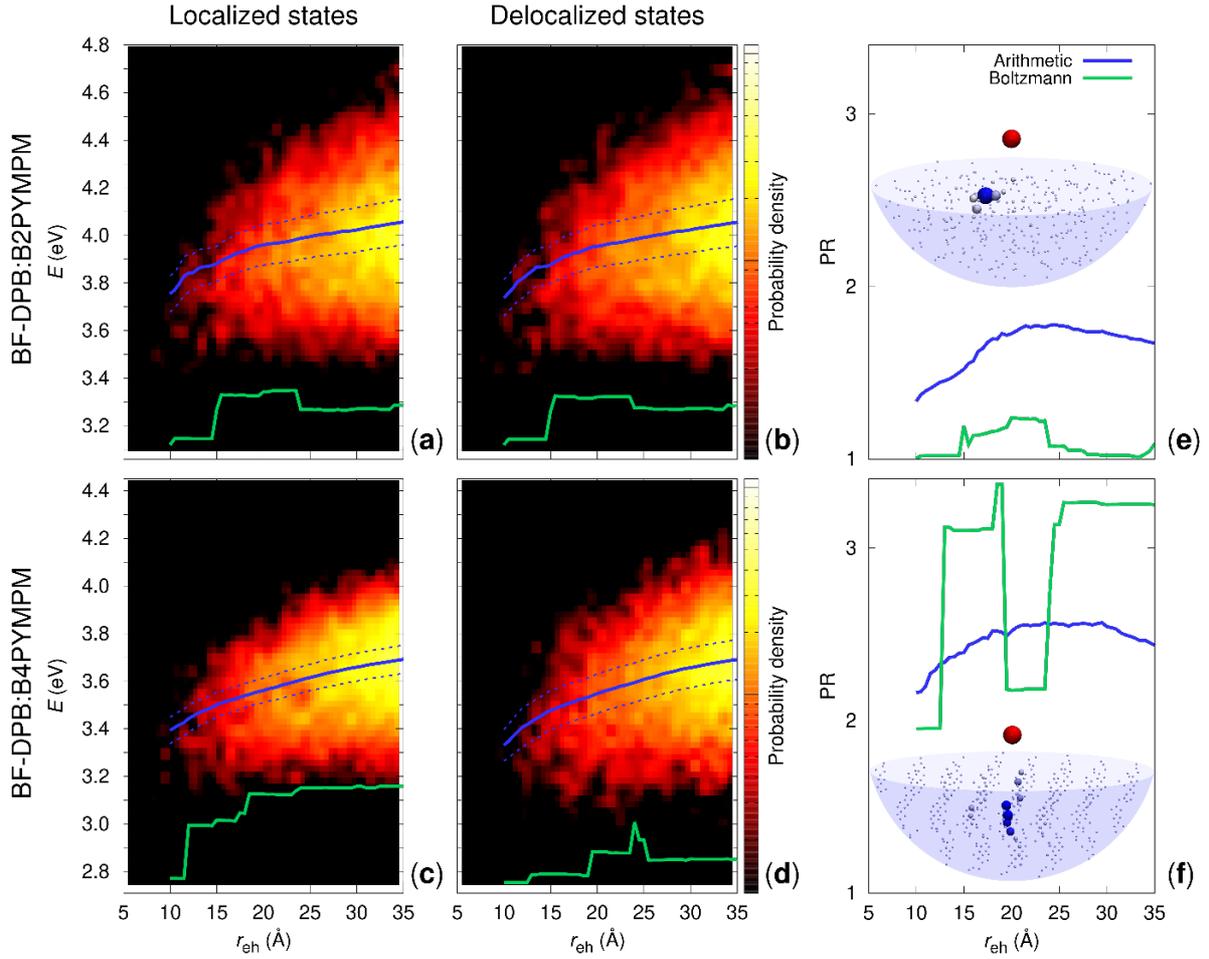

**Figure 6.** Role of charge delocalization. Evolution of the calculated CT DOS (**a-d**) and of the PR (**e,f**) as a function of the e-h distance ($r_{eh}$) for BF-DPB:B2PYMPM (**a,b**), and BF-DPB:B4PYMPM (**c,d**) interfaces. The lines in each plot represent the arithmetic average with its standard deviation (solid and dashed blue lines, respectively) and the Boltzmann average at 300 K (in green). The spike in the DOS in the Boltzmann-averaged energy profile at around 20-25 Å in panel (**d**) and the corresponding drop in PR in panel (**f**) are ascribable to a poor statistical sampling in that region. These data show that while charge delocalization is negligible in the B2PYMPM system, it plays a major role in BF-DPB:B4PYMPM, broadening the DOS and reducing the energy barrier that thermalized CT excitons must overcome to split into free charge carriers. In panels (**e,f**) it is also shown as an inset a representative delocalized state for the BF-DPB:B2PYMPM (**e**) and the BF-DPB:B4PYMPM (**f**) samples. The color scale (from white to blue) and the volume of the spheres representing the acceptor molecules are proportional to the site population (see Supporting Information).

The present theoretical analysis shows that the high charge generation efficiency of B4PYMPM-based blends is rooted in the higher structural order, which gives rise to larger transfer integrals and lower energetic disorder as compared to the B2PYMPM heterojunction. These factors[19] prompt rather extended CT states with reduced e-h pair binding energy.

# 7. Conclusion

In conclusion, we demonstrated that the fine-tuning of chemical structure can significantly affect molecular packing, which in turn has a great impact on the energetic landscape and charge delocalization properties of molecules. Computational results showed that more structural ordered B4PYMPM molecules provide more delocalized states for charge transfer, minimizing the e-h binding energy and favoring efficient charge separation. Consequently, B4PYMPM-containing systems exhibit fast, field-independent charge generation leading to FFs over 70%. In contrast, the generation of free charge carriers in B2PYMPM and B3PYMPM systems show a strong field dependency. This is due to the existence of a CT binding energy (> ~110 meV) which restrains CT exciton dissociation. Even though CT states dissociate faster into free charge carriers than they decay in the B4PYMPM system, its emission quantum efficiency is not lower than that of the B2PYMPM and B3PYMPM systems where the CT states are more strongly bound. This is encouraging, as a high radiative recombination efficiency, in combination with a high CT state dissociation yield, is required for low voltage loss, high efficiency organic photovoltaics. This work shows that an interface with such favorable properties can be achieved by an ordered interfacial morphology enabling delocalized states in the direction perpendicular to the interface.

# 8. Experimental Section/Methods

*Device preparation*: The materials were purchased by Luminescence technology corp., Taiwan (NDDP, m-MTDATA, NPB, BPAPF, B2PYMPM, B3PYMPM, B4PYMPM); abcr GmbH, Germany (BPhen); Synthon, Germany (BF-DPB). The layers of the diodes were thermally evaporated at ultra-high vacuum (base pressure < $10^{-7}$ mbar) on a glass substrate with a pre-structured ITO contact (Thin Film Devices, USA). Glass substrates were cleaned in a multi-step wet process including rinsing with N-methyl-2-pyrrolidone, ethanol, and deionized water as well as treatment with ultraviolet ozone. Details on the device structure, donor molecules, and related energy levels are shown in Figure S1, Supporting Information. All organic materials were purified 2-3 times by sublimation. The device area is defined by the geometrical overlap of the bottom and the top contact and equals 6.44 mm$^2$. To avoid exposure to ambient conditions, the organic part of the device was covered by a small glass substrate, glued on top.

*Current-voltage characteristics:* the *J-V* curves in dark and under solar illumination were measured with a SMU (Keithley 2400, USA) at room temperature in ambient conditions. The cells were illuminated with a spectrally mismatch corrected intensity of 100 mW cm$^{-2}$ (AM1.5G) provided by a sun simulator (16 S-150 V.3 Solar Light Co., USA). Masks were used to minimize edge effects and to define an exact photoactive area (2.78 mm²). The intensity was monitored with a Hamamatsu S1337 silicon photodiode (calibrated by Fraunhofer ISE Freiburg, Germany). Light-intensity-dependent *FF* measurement of the CT-OLEDs were conducted by using three 385 nm APG2C1-385-r2 UV LEDs (Roithner, Austria) in series as illumination source and a Keithley SMU 2635A to measure the current-voltage curve.

*EQE$_{PV}$ measurements*: EQE$_{PV}$ was measured using masks to minimize edge effects and to define an exact photoactive area (2.78 mm²). The EQE$_{PV}$ was detected with a lock-in amplifier (Signal Recovery SR 7265) under monochromatic illumination (Oriel Xe Arc-Lamp Apex Illuminator combined with Cornerstone 260 1/4m monochromator, Newport, USA) using a calibrated mono-crystalline silicon reference diode (Hamamatsu S1337 calibrated by Fraunhofer ISE, Germany). For sensitively measured EQE$_{PV}$ the light of a white high-power LED (LED Engin LZP-00CW00, USA), used for the high-$E_{CT}$ devices, was chopped at 140 Hz and coupled into a monochromator (Newport Cornerstone 260 1/4m, USA). The resulting monochromatic light was focused onto the OSC, its current at short-circuit conditions was fed to a current pre-amplifier before it was analyzed with a lock-in amplifier (Signal Recovery 7280 DSP, USA). The time constant of the lock-in amplifier was chosen to be 1s and the amplification of the pre-amplifier was increased to resolve low photocurrents. The EQE$_{PV}$ was determined by dividing the photocurrent of the OSC by the flux of incoming photons, which was measured using a calibrated Si and InGaAs photodiode (FDS100-CAL and FGA21-CAL, Thorlabs Inc., USA).

*Electroluminescence measurements*: The EL spectra were obtained with an Andor SR393i-B spectrometer equipped with a cooled Si and cooled InGaAs CCD detector array (DU420A-BR-DD and DU491A-1.7, UK). The spectral response of the setup was calibrated with a reference lamp (Oriel 63355). The emission spectrum of the OSCs was recorded at different injection currents with respect to voltages, which were lower than or at least similar to the $V_{OC}$ of the device at 1 sun illumination. Additional certification of the EL measurements was determined by a flux calibrated Acton SpectraPro SP2560 monochromator coupled to a cooled Spec10LN Si CCD camera from Princeton Instruments.

*EQE$_{EL}$ measurements*: The EQE$_{EL}$ was measured by forward biasing the OSCs with either an Agilent 4155C parameter analyser or Keithley SMU and collecting the emitted radiation by an enhanced G10899-03K InGaAs photodetector from Hamamatsu. The absolute total photon flux determination was performed by placing the OSC at a distance of 18.3 mm from the photodetector. Knowledge about the spectral distribution of the cell emission, the spectral response of the InGaAs photodetector, and the assumption of a point source emitting uniformly into a half-sphere allowed for the determination of the absolute EL photon flux from the OSC. Uncertainties in measured EQE$_{EL}$ are expected to be governed by the small distance imprecision between the OSC and the photodetector (calibrated Si detector from Newport, 818-series with an active area of 1 cm$^2$). To keep this uncertainty as little as possible, the measurement was conducted in different distances from the solar cell and always extrapolated to the full half sphere.

*Temperature dependent suns-$V_{OC}$ and EL measurements:* For suns-$V_{OC}$ measurements, a Keithley SMU2635A was controlling the LED (a white LED (APG2C3-NW, Roithner, Austria) for the OSCs and a 365 nm LED (APG2C1-365-r4, Roithner) for the CT-OLEDs) to change the light intensity. A Keithley dual channel SMU2602A measures both the $V_{OC}$ and the illumination intensity with a Newport 818-UV photodiode. To measure the EL, the dual channel SMU2602A applied a bias voltage to the sample, and measures the photocurrent of a S2387-66R Si Photodiode (Hamamatsu, Japan), which was directly attached to the device, covering the whole active area. To change the cell temperature, the devices were placed in vacuum on a copper block, which was connected to a Peltier element from Peltron GmbH (Fürth, Germany), controlled by a BelektroniG HAT Control device (Freital, Germany).

*TDCF measurements:* The TDCF experiment was performed by using a laser pulse from a diode pumped, Q-switched Nd:YAG laser (NT242, EKSPLA) with 6 ns pulse duration and a typical repetition rate of 500 Hz working at 404 nm to generate charges in the device. A pulse generator (Agilent 81150A) was used to apply the pre- and collection bias which are amplified by a home-built amplifier. The current through the device was measured via a grounded 10 Ω resistor in series with the sample and with a differential current probe recorded with an oscilloscope (DSO9104H). The pulse generator was triggered with a fast photodiode (EOT, ET-2030TTL). The fluence was determined with a CCD-camera in combination with a calibrated photodiode sensor (Ophir) and a laser-cut high-precision shadow mask to define the illuminated area.

*GIWAXS measurements*:

The GIWAXS measurements were performed at BL11 (NCD-SWEET) at the ALBA synchrotron in Barcelona, Spain. For the measurements a Rayonix LX255-HS detector was used approximately 140 mm behind the sample. A beam energy of 12.4 keV, exposure time of 120s and an incidence angle of 0.12° were used. The analysis of the scattering data was performed using WxDiff (© S.C.B.M.). The population analysis is performed by integrating the intensity along a Chi-Arc from 2° to 80° degrees in a specific Q range (ca. 1.4-1.9 Å$^{-1}$). The integrated intensity is corrected by cos($\chi$), due to decrease in intensity at small $\chi$ angles close to the horizon, which results from the grazing-incidence geometry.[31,32]

*Computational methods:*

MD simulations were performed using a timestep of 1 fs, 3D periodic boundary conditions and the particle mesh Ewald summation for electrostatic interactions, a cutoff of 12 Å for Lennard-Jones interactions. Interface samples have been equilibrated at 298 K and 1 atm before being used for further calculations. Molecular geometries were then extracted from the last MD configuration and used to calculate the parameters for a model TB electronic Hamiltonian represented on a diabatic basis of localized molecular sites. Site energies were obtained with a combination of many-body ev*GW*, DFT and ME calculations. Electron transfer integrals were computed at the DFT PBE0/def2-SVP level of theory. See Supporting Information for a full description of the computational details and further results.

**Supporting Information**

Supporting Information is available from the Wiley Online Library or from the author.

**Acknowledgements**

X.J., L.S. and G.L. contributed equally to this work. X.J. thanks support from the China Scholarship Council (no. 201706140127) and Graduate Academy of Technische Universität Dresden. The work in Bologna has been performed under the Project HPC-EUROPA3 (INFRAIA-2016-1-730897), with the support of the EC Research Innovation Action under the H2020 Program; in particular, G.L. gratefully acknowledges the support of the Department of Industrial Chemistry, University of Bologna and the computer resources and technical support provided by CINECA. The work in Mons has been supported by the European Union's Horizon 2020 research and innovation program under the Marie Skłodowska-Curie Grant agreement No. 722651 (SEPOMO project). Computational resources were provided by the Consortium des Équipements de Calcul Intensif (CÉCI), funded by the Fonds de la Recherche Scientifique de Belgique (F.R.S.-FNRS) under Grant No. 2.5020.11, as well as the Tier-1 supercomputer of the Fédération Wallonie-Bruxelles, infrastructure funded by the Walloon Region under Grant Agreement No. 1117545. L.S., G.L., L.M., G.D. and D.B. gratefully thank Xavier Blase for sharing the FIESTA code. D.B. is a FNRS Research Director. F.T. and S.C.B.M would like to acknowledge support by the German Excellence Initiative via the Cluster of Excellence EXC 1056 "Center for Advancing Electronics Dresden (cfaed)" and support from the German Research Foundation (DFG, MA 3342/6-1). The GIWAXS experiments were performed at BL11 NCD-SWEET beamline at ALBA Synchrotron with the collaboration of ALBA staff. We would like to thank Eduardo Solano and Marc Malfois for their assistance during the beam time.

Received: ((will be filled in by the editorial staff))
Revised: ((will be filled in by the editorial staff))
Published online: ((will be filled in by the editorial staff))

**ToC figure**

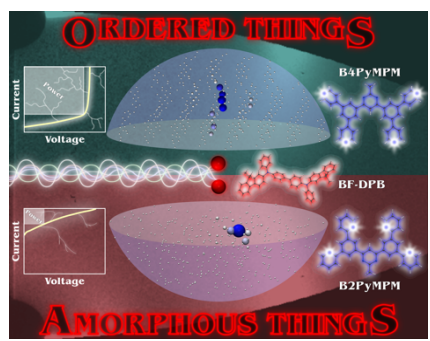